\documentclass[11pt,twoside]{article}

\usepackage{galev06}
\usepackage{epsf}
\usepackage{epsfig}
\usepackage{lscape}

\begin{document}
\title{Infrared Constrains on AGN Tori Models}   
\author{Evanthia Hatziminaoglou (1), the SWIRE Team}   
\affil{(1) Instituto de Astrof\'{i}sica de Canarias, C/ V\'{i}a L\'{a}ctea s/n, 38300 La Laguna, Spain}    

\begin{abstract} 
This work focuses on the properties of dusty tori in active galactic nuclei (AGN) derived
from the comparison of SDSS type 1 quasars with mid-Infrared (MIR) counterparts
and a new, detailed torus model. The infrared data were taken by the Spitzer Wide-area
InfraRed Extragalactic (SWIRE) Survey. Basic model parameters are constraint,
such as the density law of the graphite and silicate grains, the torus size and its
opening angle. A whole variety of optical depths is supported. The favoured models
are those with decreasing density with distance from the centre, while there is no
clear tendency as to the covering factor, i.e. small, medium and large covering factors are
almost equally distributed. Based on the models that better describe the observed SEDs,
properties such as the accretion luminosity, the mass of dust, the inner to outer radius 
ratio and the hydrogen column density are computed. The properties of the tori, as derived 
fitting the observed SEDs, are independent of the redshift, once observational biases are 
taken into account.
\end{abstract}


\section{Torus Model}   

A detailed description of the torus model is given in \cite{fritz06}; here
we briefly present the main characteristics. 
The adopted torus geometry is a {\it flared disk}, i.e.
a sphere with the polar cones removed. Its size is defined by its outer radius, 
R$_{\rm out}$, and its opening angle. The dust components that dominate both the 
absorption and the emission of radiation are
graphite and silicate. The inner radius, R$_{\rm in}$, depends both on the
sublimation temperature of the dust grains (1500 and 1000 K, for graphite and silicate,
respectively) and on the accretion luminosity. Since silicate grains have a lower 
sublimation temperature than graphite grains, the innermost regions of the torus
only consist of graphite. The adopted absorption
and scattering coefficients are those by \cite{laor93} for dust grains of different
dimensions, weighted with the standard MRN distribution \citep{mathis77}.
The gas density within the torus is modeled in such a way to allow
a gradient along both the radial and the angular coordinates.

\begin{figure}[!ht]
\includegraphics[angle=-90, width=11cm]{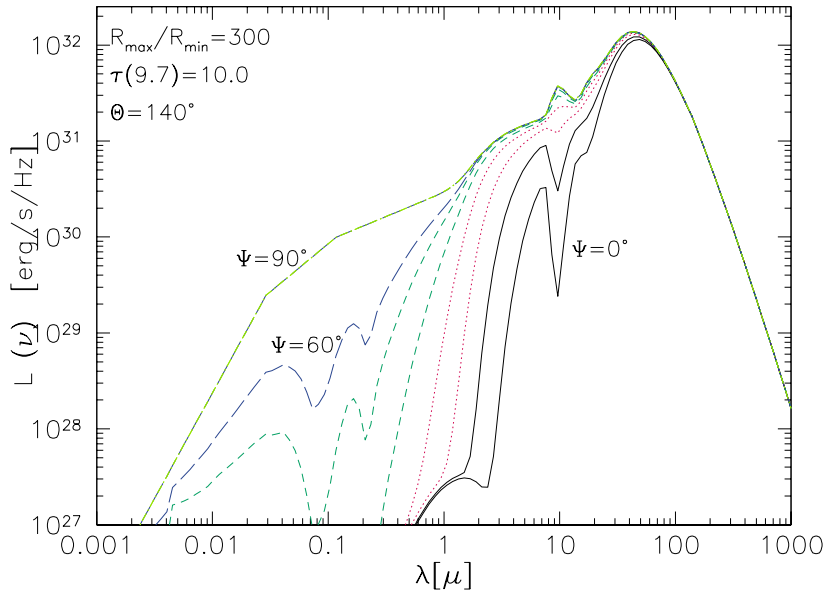}
\caption{Emission spectra as a function of wavelength for 10 different
lines-of-sight inclinations from ${\Psi} = 0^\circ$ (edge-on; lower curve) to ${\Psi} = 90^\circ$ 
(face-on, upper curve) at regular steps of $10^\circ$, for a geometrical
configuration with $\tau(9.7)=10.0$, R$_{\rm out}$/R$_{\rm in}=300$, torus
opening angle $\Theta=140^\circ$ and density law
$\rho\left(r,\theta \right) \propto r^{-1} e^{-6 |cos(\theta)|}$, where $r$ is 
the distance from the centre and $\theta$ the angle from the equatorial plane.}
\label{figMod}
\end{figure}

The central source is assumed to be point-like and its emission isotropic. Its spectral
energy distribution is defined by means of a composition of power laws with
different values for the spectra index in the UV, optical and IR.
The $\Lambda$-iteration method is used to solve the radiative transfer equation.
A geometrical grid is defined along the three spatial coordinates, and the main
physical quantities (dust density and temperatures, electromagnetic emission, optical
depth, etc.) are computed with respect to the center of the volume elements defined
by the grid. The global SED is computed at different angles of the line-of-sight with respect
to the torus' equatorial plane and includes three contributions: emission from the AGN, 
thermal emission and scattering emission by dust in each volume element. 
An example of an emitted spectrum is given in Fig. \ref{figMod}. 

\section{Data}

The MIR data used here are taken from the SWIRE EN1 and EN2 and
Lockman fields and the SWIRE catalogues processed by the
SWIRE collaboration. Details about the data can be found in \cite{lonsdale04},
\cite{surace05} and \cite{shupe06}.
The optical data were taken from the Sloan Digital Sky Survey (SDSS)
Data Release 4 (DR4), that covers the entire Lockman and SWIRE EN2
fields and a part of the SWIRE EN1 field. 2MASS data points were also
used, whenever available as well as data from the Deep 2MASS Survey in
Lockman \citep{beichman03}. 

A total of 279 spectroscopically
confirmed quasars, all detected in all IRAC bands and
MIPS 24 micron, lie within the SWIRE fields covered by the SDSS DR4
spectroscopic release, with redshifts spanning from 0.214 to 5.215.
Their $i$-band magnitudes reach 19.1 for objects with redshifts
typically less than 2.3 and go up to a magnitude deeper for higher
redshifts \citep{richards02}. For a detailed analysis of the properties of 
the sample in EN1 see \cite{eva05}.

\section{First Results and Implications}

\begin{figure}[!ht]
\plotone{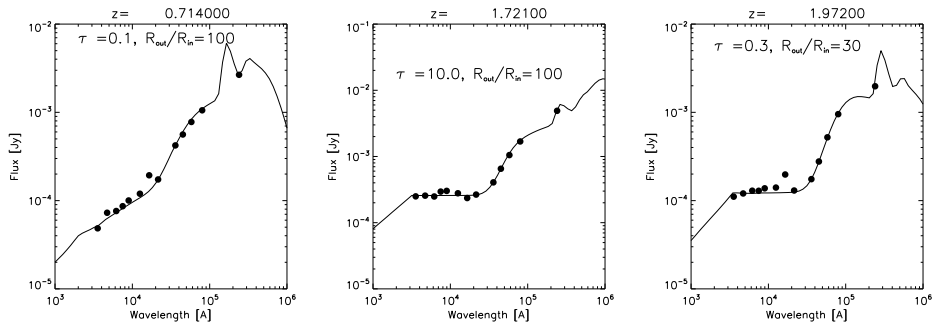}
\caption{Examples of SED fits for three AGN of our sample. The fluxes
are given in Jy, the wavelengths in \AA.}
\label{figEx}
\end{figure}

Each observed SED is compared to a total of 720 models, with 10 lines of
sight each (from an equatorial to a pole-on view). Examples of best 
fits are shown in Fig. \ref{figEx}. Some of the results of the SED fitting 
are summarised below.

\noindent
For $\sim$70\% of the objects the density, $\rho(r, \theta) \propto
r^{\beta} e^{-\gamma |cos(\theta)|}$, decreases radially from the centre 
($\beta < 0$) while for another $\sim$25\% it remains constant ($\beta = 0$). 
Half of the objects have a density
distribution that does not vary with the angle, $\theta$, from the equator.
Trying not to exclude {\it a priori} any of the model characteristics,
we allow for low optical depth tori. Some 
65\% of the objects were better described by models with $\tau_{9.7} <$ 1.0. 
We therefore need to recalculate the covering factor and consider 
that we switch from a type 1 to a type 2 object (i.e. that we
no longer see the central source through the torus) when $\tau_{0.3} > 1$.

\begin{figure} 
\plottwo{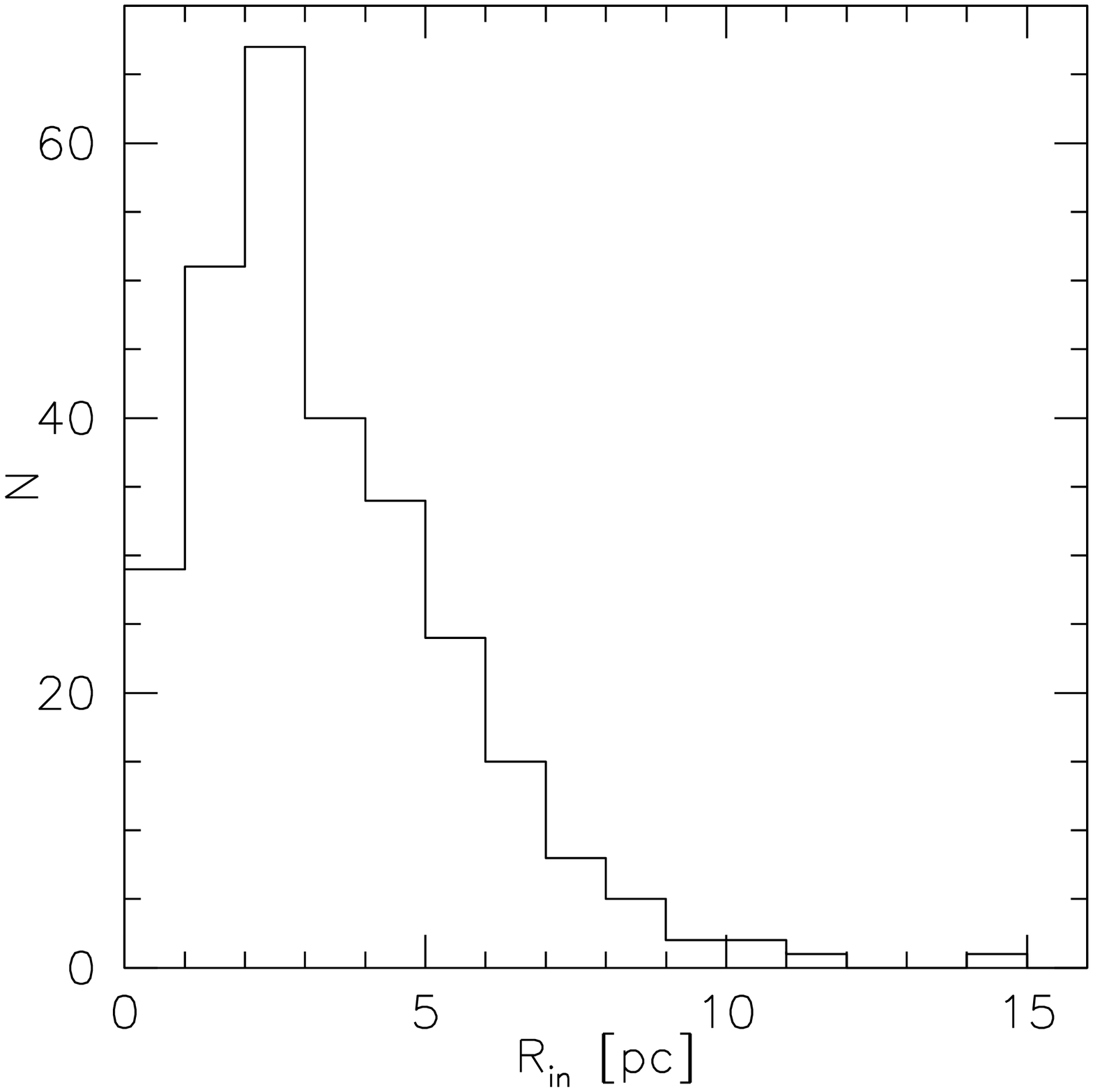}{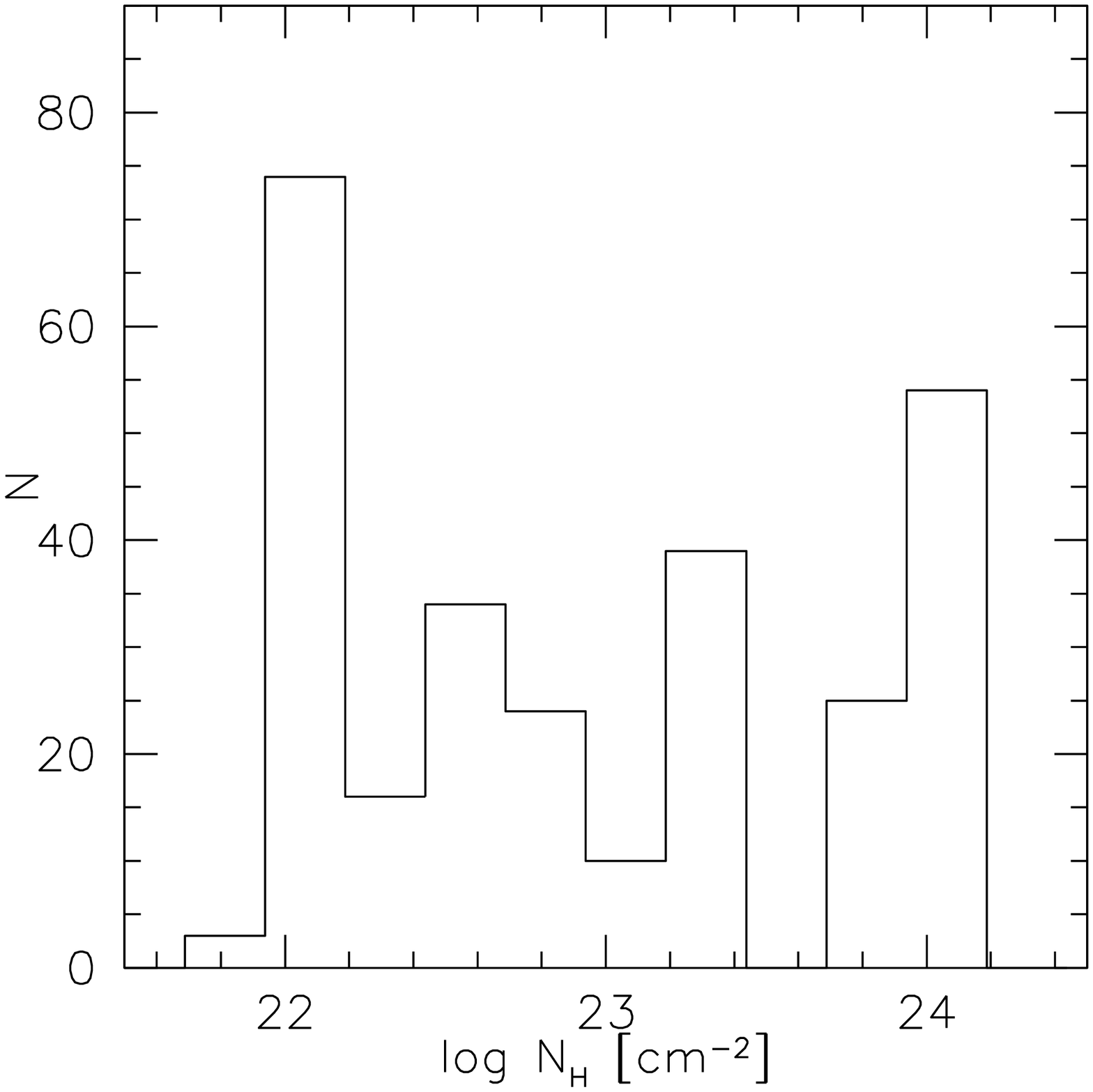}
\caption{Left panel: histogram of the inner torus radii, R$_{\rm in}$;
Right panel: distribution of the hydrogen column density, N$_{\rm H}$, along the
equator.}
\label{FigRin}
\end{figure}

Inner radii are typically of order of a few pc, with extreme cases (very
luminous objects) reaching
$\sim10$ pc, as seen on the left panel of Fig. \ref{FigRin}.
The mass of dust, M$_{\rm Dust}$, is computed summing the individual sample elements 
of the best-fit model. Typical values range from 10$^4$ to 10$^6$ M$_{\odot}$. Note 
that M$_{\rm Dust}$ does not refer to the total mass of the torus, that can be 
obtained by adding the mass of gas, typically $\sim$100 times larger than that of
the dust. Many quantities (R$_{\rm in}$, R$_{\rm out}$, M$_{\rm Dust}$,) show a clearly
increasing trend with redshift, $z$. This could be interpreted as an indication for larger
and/or more massive tori at higher $z$; however it is simply the manifestation
of the Scott effect, i.e. the fact that in order for distant objects to make it into a
flux-limited sample they need to be intrinsically brighter. What happens in reality is
that the inner radius, R$_{\rm in}$, directly scales with the accretion luminosity, 
L$_{\rm acc}$ (expressed in units of 10$^{46}$ erg/sec), via the relation by \cite{barvainis87}: 
R$_{\rm in} \simeq 1.3 \times \sqrt(\rm L_{\rm acc}) \times T_{1500}^{-2.8}$ [pc], where the 
temperature, T, is given in units of 1500K. In a flux-limited sample L$_{\rm acc}$, which
is directly related to the bolometric luminosity, increases with $z$ and therefore
so does R$_{\rm in}$, as well as R$_{\rm out}$ and M$_{\rm Dust}$ that directly depend on 
R$_{\rm in}$, as already mentioned. We therefore deduce that sources of the same 
accretion luminosity can have similar properties, independently on the redshift.

The {\it equatorial}  hydrogen column density, N$_{\rm H}$, distribution, derived 
from the models, is shown in the right panel of Fig. \ref{FigRin}. The values are 
representative of type 2 objects, with some in the range of Compton thick AGN 
(N$_{\rm H} \sim 10^{24}$ cm$^{-2}$). If the distinction between type 1 and type 2 objects
is simply a matter of orientation, as suggested from the Unified Scheme, this would
be the expected distribution of {\it equatorial}  N$_{\rm H}$.

\section{Conclusions}
This work in progress provides useful insights into the properties of dusty tori around
AGN. Fitting the observed optical-to-MIR SEDs to a series of models, we showed that
in most of the cases the density decreases with distance from the centre and can vary
with the angle from the equator. Tori properties such as the outer radii or the mass of
dust are independent of the redshift and only depend on the luminosity of the objects.
The {\it equatorial}  hydrogen column density distribution is in agreement with what is
expected from the Unified Scheme. 

\acknowledgements 
This work is based on observations made with the {\it Spitzer Space Telescope}.
It also makes use of the SDSS and 2MASS Archives. 
It was supported in part by the Spanish Ministerio de
Educaci\'{o}n y Ciencia (Grant Nr. ESP2004-06870-C02-01).

\end{document}